\documentclass[
superscriptaddress,reprint,aps,pra,floatfix,amsmath,amssymb
]{revtex4-2}

\makeatletter
\newcommand\thefontsize[1]{{#1 The current font size is: \f@size pt\par}}
\makeatother

\usepackage{tikz}

\usepackage{braket}
\usepackage{nicefrac}
\usepackage{ifluatex}
\usepackage[labelfont=bf, font=sf, justification=raggedright,format=plain]{caption}
\usepackage{isotope}
\ifluatex
\usepackage[detect-all]{siunitx}
\else
\usepackage[detect-all]{siunitx-v2}
\DeclareUnicodeCharacter{03BC}{µ}
\fi

\usepackage{lmodern}
\usepackage{bm}

\usepackage[colorlinks,linkcolor=blue,citecolor=blue,urlcolor=blue]{hyperref}
\usepackage{sansmath}
\usepackage{jabbrv} 

\newcommand{\bluef}{black}
\newcommand{\redf}{black}

\makeatletter
\newcommand{\mysansmath}{%
	\sffamily%
	\sansmath%
	\def\unboldmath{%
		\@nomath\unboldmath
		\sansmath%
	}%
}
\makeatother

\newcommand*{\figref}[2][]{%
	\hyperref[{fig:#2}]{%
		Fig.~\ref*{fig:#2}%
		\ifx\\#1\\%
		\else
		(#1)%
		\fi
	}%
}

\newcommand*{\figureref}[2][]{%
	\hyperref[{fig:#2}]{%
		Figure~\ref*{fig:#2}%
		\ifx\\#1\\%
		\else
		(#1)%
		\fi
	}%
}

\newcommand*{\tblref}[2][]{%
	\hyperref[{tbl:#2}]{%
		Tbl.~\ref*{tbl:#2}%
		\ifx\\#1\\%
		\else
		\,(#1)%
		\fi
	}%
}

\renewcommand{\vec}{\bm}

\newcommand*{\hfsMState}[6]{{\mathrm{#1}#2}_{\nicefrac{#3}{#4}},F=#5,m_F=#6}
\newcommand*{\hfsState}[5]{{\mathrm{#1}#2}_{\nicefrac{#3}{#4}},F=#5}

\newcommand{\vStateGroundStateHighMagneticFieldSensor}{\ket{\hfsMState{5}{S}{1}{2}{3}{-1}}}

\newcommand{\vStateGroundStateHighMagneticFieldSensorShortForm}{\ket{\uparrow}}
\newcommand{\vStateGroundStateLowMagneticFieldSensor}{\ket{\hfsMState{5}{S}{1}{2}{2}{-1}}}
\newcommand{\vStateGroundStateLowMagneticFieldSensorShortForm}{\ket{\downarrow}}
\newcommand{\vMagneticFieldSensorDetuningIntermediateLevel}{-2\pi\times\SI{9}{\giga\hertz}}

\newcommand{\sRamseyFrequency}{\omega_\mathrm{R}}
\newcommand{\sRamseyTimeOfFreePrecession}{T}

\newcommand{\sMagneticFieldSensorEffectiveDetuningGroundState}{\delta_\mathrm{eff}}

\newcommand{\sMagneticFieldSensorMagneticQuantizationField}{\vec{B}_\mathrm{q}}
\newcommand{\sMagneticFieldSensorMagneticGradientField}{\vec{B}_\mathrm{t}}

\newcommand{\sBeamRadius}{w}

\newcommand{\sRabiFrequency}{\Omega_\mathrm{R}}

\newcommand{\sArrayTrapDepth}{U}

\newcommand{\sMF}{m_\mathrm{F}}

\newcommand{\vMagneticFieldSensorTypicalArrayTrapDepth}{k_\mathrm{B}\times\SI{1}{\milli\kelvin}}

\newcommand{\vMagneticFieldSensorDetectionExposureTime}{\SI{60}{\milli\second}}
\newcommand{\vMagneticFieldSensorQuantizationFieldStrength}{\SI{283(1)}{\micro\tesla}}

\newcommand{\vMagneticFieldSensorTwoPhotonRabiFrequency}{2\pi\times\SI{0.6(1)}{\mega\hertz}}

\newcommand{\vMagneticFieldSensorReferenceFrequency}{2\pi\times\SI{38.7(13)}{\kilo\hertz}}

\newcommand{\vMagneticFieldSensorGradientMean}{\SI{76.9(7)}{\nano\tesla/\micro\meter}}

\newcommand{\sMagneticFieldSensorReferenceFrequency}{\omega_\mathrm{R,ref}}

\begin{document}
\title{Quantum Sensing in Tweezer Arrays: \\Optical Magnetometry on an Individual-Atom Sensor Grid}
\author{Dominik Sch\"{a}ffner}
\affiliation{Technische Universit\"at Darmstadt, Institut f\"ur Angewandte Physik, Schlossgartenstra\ss e 7, 64289 Darmstadt, Germany}
\author{Tobias Schreiber}
\affiliation{Technische Universit\"at Darmstadt, Institut f\"ur Angewandte Physik, Schlossgartenstra\ss e 7, 64289 Darmstadt, Germany}
\author{Fabian Lenz}
\affiliation{Technische Universit\"at Darmstadt, Institut f\"ur Angewandte Physik, Schlossgartenstra\ss e 7, 64289 Darmstadt, Germany}
\author{Malte Schlosser}
\affiliation{Technische Universit\"at Darmstadt, Institut f\"ur Angewandte Physik, Schlossgartenstra\ss e 7, 64289 Darmstadt, Germany}
\author{Gerhard Birkl}
\email[]{Contact for correspondence: apqpub@physik.tu-darmstadt.de}
\homepage[]{https://www.iap.tu-darmstadt.de/apq}
\affiliation{Technische Universit\"at Darmstadt, Institut f\"ur Angewandte Physik, Schlossgartenstra\ss e 7, 64289 Darmstadt, Germany}
\affiliation{Helmholtz Forschungsakademie Hessen f\"ur FAIR (HFHF), Campus Darmstadt, Schlossgartenstra\ss e 2,
	64289 Darmstadt, Germany}
\date{\today}

\begin{abstract}		
	We \textcolor{\bluef}{implement} a scalable platform for quantum sensing comprising hundreds of sites capable of holding individual laser-cooled atoms and demonstrate the applicability of this single-quantum-system sensor array to magnetic-field mapping on a two-dimensional grid. With each atom being confined in an optical tweezer within an area of \SI{0.5}{\micro\meter\squared} at mutual separations of \SI{7.0(2)}{\micro\meter}, we obtain micrometer-scale spatial resolution and highly parallelized operation. An additional steerable optical tweezer allows for a rearrangement of atoms within the grid and enables single-atom scanning microscopy with sub-micron resolution. \textcolor{\bluef}{This} individual-atom sensor platform finds its immediate application in mapping an externally applied DC gradient magnetic field. In a Ramsey-type measurement, we obtain a field resolution of \SI{98(29)}{\nano\tesla}. \textcolor{\bluef}{We estimate the sensitivity to $\SI{25}{\micro\tesla/\sqrt{\hertz}}$}.\\
		
	\noindent
	\textcolor{blue}{
		Published as: PRX Quantum \textbf{5}, 010311 (2024), \url{https://doi.org/10.1103/PRXQuantum.5.010311}
	}
	\end{abstract} 
\maketitle

\section{introduction}

The paradigm of exploiting quantum properties of physical systems for sensing purposes has paved the way to devices with unprecedented sensitivity that outshine their classical counterparts \cite{Degen2017,Mitchell2020}.
The fact that energy levels of quantum systems in general are susceptible to their environment allows the measuring of, for example external electric or magnetic fields. In this context, devices based on superconducting circuits \cite{Cohen1972,Vasyukov2013,Danilin2018,Stolz2022}, nitrogen-vacancy centers
\cite{Maletinsky2012,Rondin2014,Barry2020,Scholten2021,Zhang2022},
and optically pumped atomic magnetometers \cite{Budker2007,Kitching2018,Amico2021} have been developed to highly sensitive systems for magnetic field sensing. Each approach exhibits a specific set of advantages and limitations due to its characteristics derived from the physical properties when benchmarked with respect to sensitivity, spatial resolution, field of view, or ambient conditions---to name only a few crucial parameters.\\
In recent years, this technological toolbox has been enriched by methods based on ultracold-neutral-atom quantum systems. Localized atom clouds represent well-controllable sensors combining high-precision measurements with micrometer-scale spatial resolution and millimeter-scale measurement regions.
Building on miniaturized magnetic traps, ultracold-atom magnetic --field microscopy renders the simultaneous observation of microscopic and macroscopic effects possible \cite{Aigner2008,Yang2017}. Optical dipole traps establish a basis for sensors that are detached from any perturbing surface, thereby facilitating high-precision all-optical magnetometers \cite{Vengalattore2007,Dang2010,Eto2013,Eliasson2019} whose sensitivity has been further amplified by entanglement-enhanced measurements \cite{Sewell2012,Muessel2014}. In addition to magnetic field mapping, further applications include the rectification of the magnetic field environment for neutral-atom quantum science. This has been demonstrated by counteracting on global fluctuations in Bose-Einstein condensates \cite{Eto2013} and error mitigation in quantum simulation and computing \cite{Eliasson2019,Singh2023}.\\
\textcolor{\bluef}{
In particular, optical tweezer arrays \cite{Kaufman2021} facilitate spatially configurable quantum sensing devices of inherently identical single quantum systems that are candidates for surpassing the energy resolution limit \cite{Mitchell2020}. Because of the free-space realization of fully isolated single-spin systems far away from any surface, this sensor platform avoids detrimental effects of surface noise, interaction-induced energy shifts, and depolarization effects \cite{Mitchell2020b}.
In addition, building on the technology of cold-atom frequency standards and optical array clocks \cite{Norcia2019,Madjarov2019}, and being further fostered by
the exploitation of tailored quantum states and the application of elaborate quantum gate sequences \cite{Montenegro2022,Singh2023,Hines2023,Eckner2023,Bornet2023}, this platform has the potential of reaching the ultimate limit in resolution and accuracy and defining the frontier of optimal quantum sensing \cite{Marciniak2022}.
Together with the prospect of transferring the atoms into Rydberg states with additional susceptibility to external fields \cite{Hines2023,Eckner2023,Bornet2023}, this platform exhibits excellent prospects for massively parallelized two-dimensional (2D) and three-dimensional (3D) \cite{Barredo2018,Wu2019,Schlosser2023} sensing of magnetic, electric, and radio-frequency fields.}\\
We report on the realization of an individual-atom magnetic field sensor grid in a highly scalable architecture.
Individual neutral atoms are loaded into microlens-generated tweezers \cite{Dumke2002,OhldeMello2019} with micrometer-sized pitch for the measurement of the ambient magnetic field. We use this sensor array to map an externally applied magnetic gradient field. An auxiliary transport tweezer is used to increase the duty cycle by maintaining a high atom number within a subarray for local measurements. In addition, this steerable tweezer increases the spatial resolution as an individual probe decoupled from the array grid. While already providing hundreds of sensor pixels in two dimensions in this work, our platform is intrinsically scalable to 3D arrays with thousands of sensor atoms as presented in Ref.~\cite{Schlosser2023}.

\section{experimental details}
We realize magnetic field sensing in the neutral-atom platform described in detail in Refs.~\cite{OhldeMello2019,Schlosser2020} and schematically illustrated in \figref[a]{opticalSetup}.
Created by microfabricated optical elements, this setup constitutes a two-dimensional focused-beam optical dipole trap array, i.e., tweezer array, for the parallelized, site-resolved control of single-atom quantum systems.
The tweezer spacing is \SI{7.0(2)}{\micro\meter} and the tweezer waist is $\SI{1.45(10)}{\micro\metre}$.
Throughout this article, uncertainties are given as $1\sigma$ statistical uncertainties.
The tweezers are loaded with laser-cooled \isotope[85]{Rb} atoms.
Because of light-assisted collisions, each site is either empty or occupied by one atom at most (see Refs.~\cite{Kaufman2021,Pause2023} and references therein), resulting in a typical loading efficiency of $\SI{50}{\percent}$.
From the known values of potential depth ($U=k_B\times\SI{0.2}{\milli\kelvin}$) and temperature (\SI{52}{\micro\kelvin}), we infer a 2D spatial localization of each atom of \SI{0.5}{\micro\meter\squared}.
Covering an area of ${18 \times 15}$ sites ($\SI{119}{\micro\meter}\times\SI{98}{\micro\meter}$), this configuration serves as planar sensor grid for magnetic field mapping. Our setup includes an additional movable optical tweezer with a waist of $\SI{2.0(1)}{\micro\meter}$ that can be used to arrange the randomly loaded atoms in predefined geometries.
\begin{figure}[t]
	\mysansmath
	\begin{tikzpicture}
	\node[inner sep=0pt,anchor=west] (OpticalSetup) at (0,0)
	{\includegraphics[width=\linewidth]{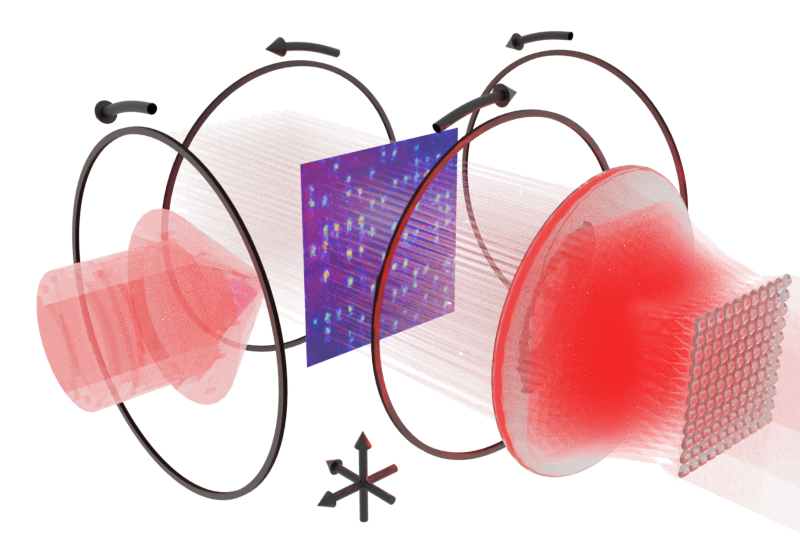}};
	
	\node[inner sep=0pt,anchor=west] (MeanImg) at (-0.1,-4.75)
	{\includegraphics[width=0.33\linewidth]{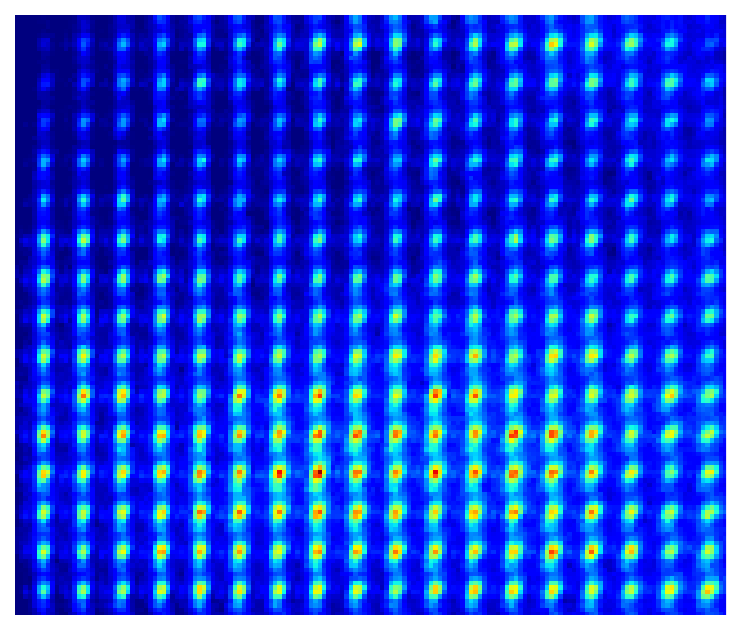}};
	
	\node[inner sep=0pt,anchor=west] (MeanImg) at (2.8,-4.75) 
	{\includegraphics[width=0.33\linewidth]{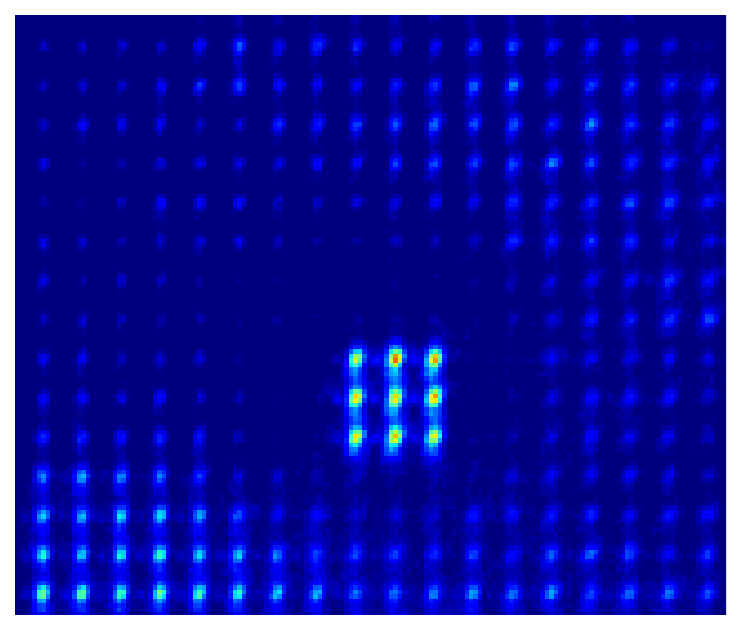}};
	
	\node[inner sep=0pt,anchor=west] (MeanImg) at (5.7,-4.75)
	{\includegraphics[width=0.33\linewidth]{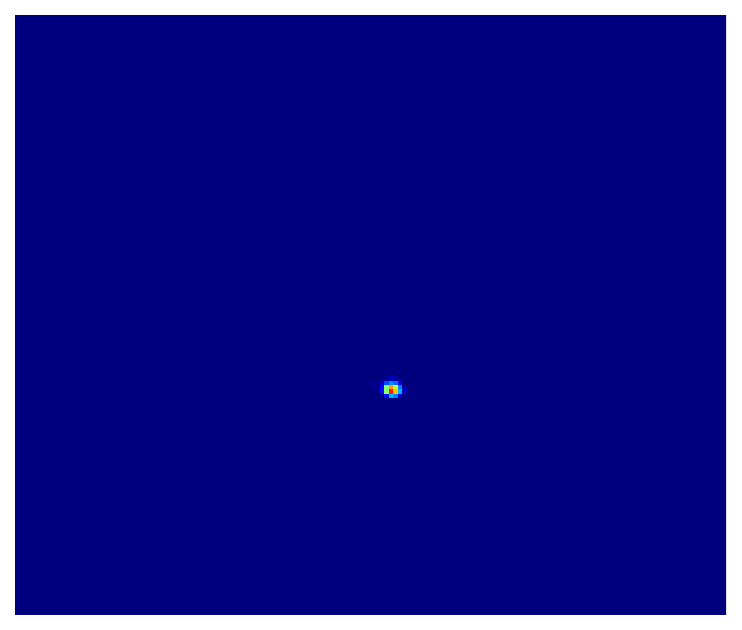}};
	\node[anchor=west,inner sep=0] at (-0.1,2.8) {\textbf{(a)}};	
	\node[anchor=west,inner sep=0] at (-0.1,-3.3) {\textbf{(b)}};
	\node[anchor=west,inner sep=0] at (2.8,-3.3) {\textbf{(c)}};
	\node[anchor=west,inner sep=0] at (5.7,-3.3) {\textbf{(d)}};
	
	\node[anchor=center] at (7.6,-2.5) {Microlens array};	
	\node[anchor=center] at (5.6,-2.6) {\centering\parbox{1.5cm}{Reimaging optics}};	
	\node[anchor=center,fill=white,fill opacity=0.8] at (1,-2.0) {\centering\parbox{2cm}{Spectroscopy beam}};
	\node[anchor=center] at (1,-2.0) {\centering\parbox{2cm}{Spectroscopy beam}};
	\node[anchor=center] at (3.5,1.65) {\centering\parbox{1cm}{Atom plane}};	
	
	\node[anchor=mid] at (3.4,2.7) {$I_\mathrm{t}$};
	\node[anchor=mid] at (5,2) {$I_\mathrm{t}$};
	\node[anchor=mid] at (1.2,2) {$I_\mathrm{q}$};
	\node[anchor=mid] at (5.9,2.8) {$I_\mathrm{q}$};
	
	\node[anchor=mid] at (3.4,-2.0) {z};
	\node[anchor=mid] at (3.9,-1.6) {y};
	\node[anchor=mid] at (3.3,-2.6) {x};
	\end{tikzpicture}
	\caption{
		Micro-optical tweezer array for cold-atom magnetometry.
		(a) Experimental setup.
		\textcolor{black}{A 270-site register of individual rubidium atoms at a wavelength of $\SI{797}{\nano\meter}$ is generated by a microlens array whose focal plane is reimaged into the vacuum chamber.}
		Two-photon Ramsey spectroscopy is used to characterize the magnetic gradient test field induced by the two anti-Helmholtz coils with counterpropagating current $I_\mathrm{t}$. A pair of coils in the Helmholtz configuration provides a well-defined quantization axis via a homogeneous field generated by a current $I_\mathrm{q}$.
		(b) Averaged fluorescence of atoms in the $\SI{18}{}\times\SI{15}{}$ tweezer array.
		(c) Enhanced atom occupation in a predefined substructure with use of a steerable optical tweezer for atom transport.
		(d) With use solely of the steerable tweezer, atoms can be positioned independently of the array grid.
		}
	\label{fig:opticalSetup}
\end{figure}\\
Site-resolved fluorescence detection of the atoms is realized through excitation at the rubidium D2 line and imaging of the atom plane onto an electron-multiplying charge-coupled device camera within an exposure time of \vMagneticFieldSensorDetectionExposureTime. \figureref[b]{opticalSetup} shows an averaged fluorescence image of atoms in the full sensor region while in \figref[c]{opticalSetup} the enhanced atom occupation in a $\SI{3}{}\times\SI{3}{}$ predefined target structure after rearrangement is depicted.
\figref[d]{opticalSetup} shows the fluorescence of single atoms loaded into the steerable optical tweezer.
\begin{figure}[b]
	\begin{tikzpicture}
	\node[inner sep=0pt,anchor=west] (MeasurementCycle) at (0,5.5) 
	{\includegraphics{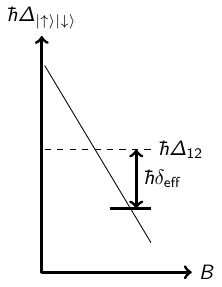}};
	\node[inner sep=0pt,anchor=west] (MeasurementCycle) at (4.2,5.5) 
	{\includegraphics{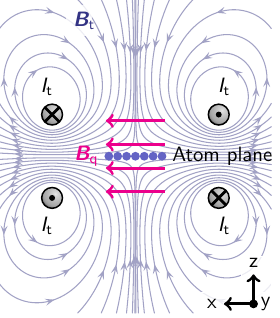}};
	\node[inner sep=0pt,anchor=west] (MeasurementCycle) at (0,0.4)
	{\includegraphics{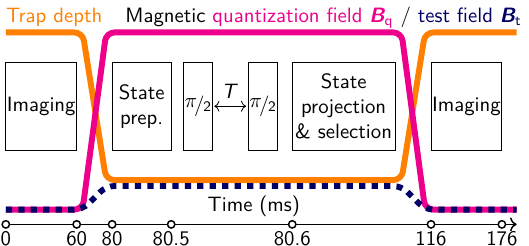}};
	\mysansmath
	\node[anchor=west] at (0,8.2) {\textbf{(a)}};
	\node[anchor=west] at (4.2,8.2) {\textbf{(b)}};
	\node[anchor=west] at (0,2.7) {\textbf{(c)}};
	\end{tikzpicture}
	\caption{
		Experimental details of individual-atom magnetometry. (a) Energy splitting $\hbar\Delta_{\vStateGroundStateHighMagneticFieldSensorShortForm\vStateGroundStateLowMagneticFieldSensorShortForm}$ of the two states $\vStateGroundStateHighMagneticFieldSensorShortForm$ and $\vStateGroundStateLowMagneticFieldSensorShortForm$ used for magnetic field sensing as a function of the magnetic field strength. For a fixed differential two-photon energy $\hbar\Delta_{12}$ of the spectroscopy beam (dashed horizontal line), the effective detuning $\sMagneticFieldSensorEffectiveDetuningGroundState$ depends on the magnetic field experienced by the atom.		
		(b) The inhomogeneous magnetic field $\sMagneticFieldSensorMagneticGradientField$ generated by a pair of coils in the anti-Helmholtz configuration is used as a position-dependent test field. A homogeneous magnetic field $\sMagneticFieldSensorMagneticQuantizationField$ defines the quantization axis.
		(c) A typical measurement cycle (see the main text for details\textcolor{\bluef}{; the time axis is not to scale}). During the Ramsey sequence, consisting of state preparation, a $\pi/2$ pulse, free precession, a $\pi/2$ pulse, and state projection and selection, the trap depth $\sArrayTrapDepth$ is $k_\mathrm{B}\times\SI{0.2}{\milli\kelvin}$ whereas for initial atom loading and the two imaging sequences  $\sArrayTrapDepth=\vMagneticFieldSensorTypicalArrayTrapDepth$.} \label{fig:levelSchemeAndMeasurementCycle}
\end{figure}\\

The spectroscopy beam in \figref[a]{opticalSetup} drives a two-photon transition between the long-lived hyperfine ground states $\vStateGroundStateHighMagneticFieldSensorShortForm=\vStateGroundStateHighMagneticFieldSensor$ and $\vStateGroundStateLowMagneticFieldSensorShortForm=\vStateGroundStateLowMagneticFieldSensor$.
This necessitates two phase-coherent frequency components, which are provided by a system of two phase-locked diode lasers \cite{Preuschoff2022}.
The laser fields are detuned by $\vMagneticFieldSensorDetuningIntermediateLevel$ relative to the single-photon transition to
$\ket{\hfsState{5}{P}{3}{2}{4}}$, while the frequency difference $\Delta_{12}$ of the two components can be precisely set in the vicinity of the field-dependent transition frequency $\Delta_{\vStateGroundStateHighMagneticFieldSensorShortForm\vStateGroundStateLowMagneticFieldSensorShortForm}$ by a stable radio-frequency source.
\figureref[a]{levelSchemeAndMeasurementCycle} shows the magnetic field dependence of the energy difference  $\hbar\Delta_{\vStateGroundStateHighMagneticFieldSensorShortForm\vStateGroundStateLowMagneticFieldSensorShortForm}$ between both hyperfine states.
Note that light shifts of the tweezer array and the spectroscopy beam modify  $\Delta_{\vStateGroundStateHighMagneticFieldSensorShortForm\vStateGroundStateLowMagneticFieldSensorShortForm}$ \cite{Ramola2021}.
%
The effective two-photon detuning is given by $\sMagneticFieldSensorEffectiveDetuningGroundState= \Delta_{12}-\Delta_{\vStateGroundStateHighMagneticFieldSensorShortForm\vStateGroundStateLowMagneticFieldSensorShortForm}$.
For a beam radius $\sBeamRadius$ of $\SI{170(20)}{\micro\meter}$ ($1/e^2$ waist) at the location of the array, we measure an on-resonance Rabi frequency ($\sMagneticFieldSensorEffectiveDetuningGroundState=0$) $\sRabiFrequency$ of $\vMagneticFieldSensorTwoPhotonRabiFrequency$ over the ${18 \times 15}$ sites.\\
%
\figureref[a]{opticalSetup} and \figref[b]{levelSchemeAndMeasurementCycle} depict the magnetic field geometry used for optical magnetometry in our individual-atom sensor grid. We apply a static homogeneous magnetic quantization field $\left|\sMagneticFieldSensorMagneticQuantizationField\right|=\vMagneticFieldSensorQuantizationFieldStrength$ induced by a pair of coils in the Helmholtz configuration along the x axis. This lifts the degeneracy of the magnetic hyperfine states and creates a well-defined quantization axis. Additionally, a pair of coils in the anti-Helmholtz configuration allows the creation of a test gradient field $\sMagneticFieldSensorMagneticGradientField$.\\
A typical measurement cycle is visualized in \figref[c]{levelSchemeAndMeasurementCycle}.
First, the occupation of atoms in the sensor grid is imaged. After activation of the quantization field $\sMagneticFieldSensorMagneticQuantizationField$ (pink), the state of the atoms is prepared in the measurement basis $\{\vStateGroundStateHighMagneticFieldSensorShortForm$,$\vStateGroundStateLowMagneticFieldSensorShortForm$\}.
Without state preparation, we find almost all of the atoms in the state $\ket{\hfsMState{5}{S}{1}{2}{3}{-3}}$ which is not accessible 
\textcolor{\bluef}{to our Ramsey sequence with $\Delta m_F = 0$ due to the direction and the polarization of the spectroscopy beam and the orientation of the quantization axis.}
Via interrupted optical pumping using $\pi$-polarized laser light at the $\ket{\hfsState{5}{S}{1}{2}{3}}\to\ket{\hfsState{5}{P}{3}{2}{3}}$ transition in combination with repumping light at the $\ket{\hfsState{5}{S}{1}{2}{2}}\to\ket{\hfsState{5}{P}{3}{2}{3}}$ transition, we can transfer about \SI{30}{\%} of the atoms to $\vStateGroundStateHighMagneticFieldSensorShortForm$.
After state preparation, no atoms remain in $\vStateGroundStateLowMagneticFieldSensorShortForm$ or other states of the $\ket{\hfsState{5}{S}{1}{2}{2}}$ manifold.
Next, a Ramsey sequence with a $\nicefrac{\pi}{2}$ pulse of duration $\tau_{\nicefrac{\pi}{2}}=\SI{0.42}{\micro\second}$ followed by a variable time $\sRamseyTimeOfFreePrecession$ of free precession is initiated.
After the second $\nicefrac{\pi}{2}$ pulse, atoms in $\vStateGroundStateHighMagneticFieldSensorShortForm$ or any other states of the $\ket{\hfsState{5}{S}{1}{2}{3}}$ manifold are removed from the tweezer array and the magnetic field is switched off again.
Finally, the remaining atoms in $\vStateGroundStateLowMagneticFieldSensorShortForm$ are detected by fluorescence imaging.
%
%
%
%
\section{magnetic field mapping}

In a sequence of measurement cycles, the periodic Ramsey signal is recorded
\textcolor{\bluef}{for 55 equidistant values of $\sRamseyTimeOfFreePrecession$ within the interval of $\SI{2}{\micro\second}\leq\sRamseyTimeOfFreePrecession\leq \SI{110}{\micro\second}$}
for a fixed differential two-photon energy $\hbar\Delta_{12}$ of the Ramsey spectroscopy beam (dashed horizontal line in \figref[a]{levelSchemeAndMeasurementCycle}).
\textcolor{\bluef}{A typical signal obtained for one of the central sensor grid positions is displayed in \figref[a]{RamseyAllan}. We fit the data with an oscillatory function that includes dephasing-induced damping (red curve).
%
Analyzing all sensor grid positions, we observe a mean time constant for $e^{-1}$ amplitude damping of $T_2^*=\SI{118(33)}{\micro\second}$.}
The \textcolor{\bluef}{extracted} Ramsey frequency $\sRamseyFrequency$, given as $2\pi$ divided by the period of the Ramsey \textcolor{\bluef}{oscillations}, directly corresponds to the effective detuning $\sMagneticFieldSensorEffectiveDetuningGroundState=\Delta_{12}-\Delta_{\vStateGroundStateHighMagneticFieldSensorShortForm\vStateGroundStateLowMagneticFieldSensorShortForm}=\sRamseyFrequency$.
Accordingly, the differential shift of the basis states at the position of a sensor atom is revealed and can be translated into the magnetic field strength with use of the magnetic field dependence of the effective detuning of $\frac{\partial \sMagneticFieldSensorEffectiveDetuningGroundState}{\partial B}=2\pi\times\SI{9.2777(3)}{\kilo\hertz/\micro\tesla}$.\\
In a first application, we use the 2D sensor grid to map the applied gradient test field $\sMagneticFieldSensorMagneticGradientField$ depicted in \figref[a]{opticalSetup}, \figref[b]{levelSchemeAndMeasurementCycle}, and \figref[c]{levelSchemeAndMeasurementCycle} (dashed line).
\begin{figure}[b]
	\begin{tikzpicture}
	\node[inner sep=0pt,anchor=west] (RamseyOscill) at (0,5.5) 
	{\includegraphics[width=0.47\linewidth]{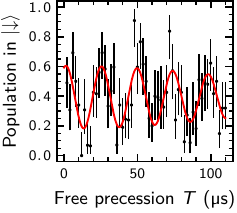}};
	\node[inner sep=0pt,anchor=west] (AllanDev) at (4.2,5.5) 
	{\includegraphics[width=0.525\linewidth]{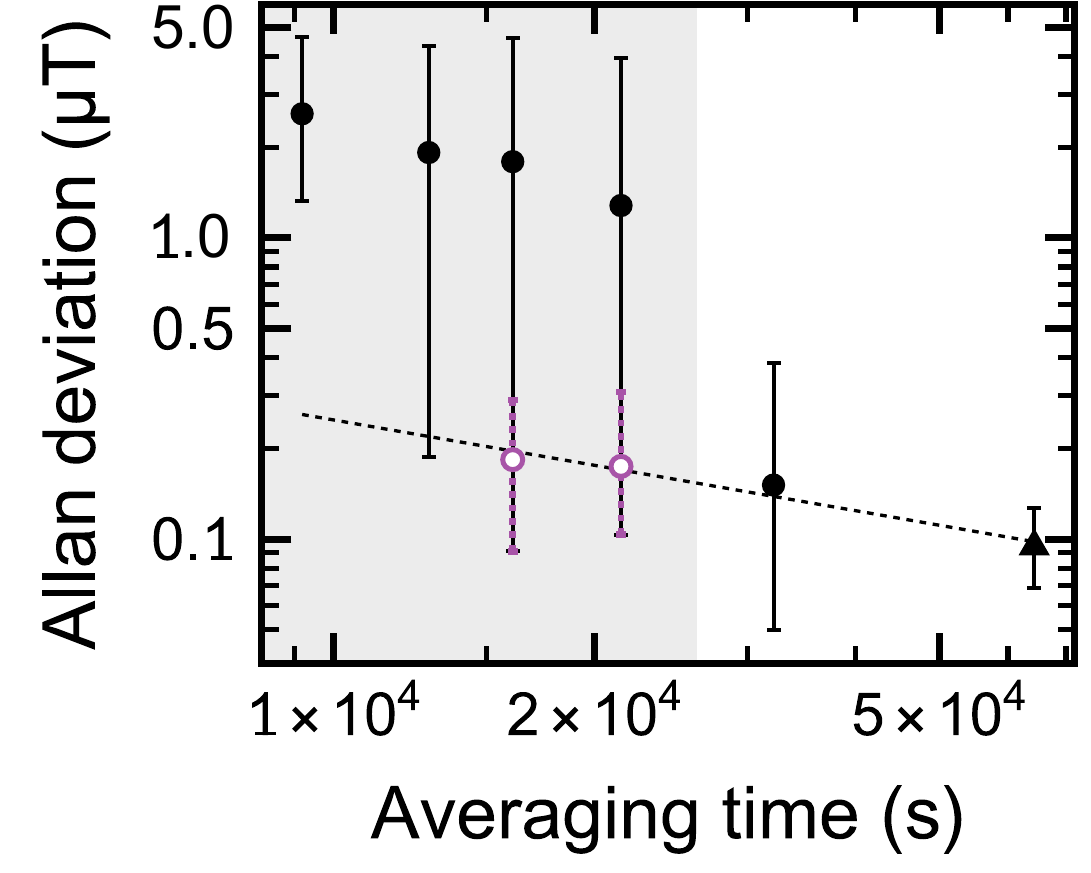}};
	\mysansmath
	\node[inner sep=0pt,anchor=west] at (0,7.7) {\textbf{(a)}};
	\node[inner sep=0pt,anchor=west] at (4.3,7.7) {\textbf{(b)}};
	\end{tikzpicture}
	\caption{(a) Ramsey signal for one of the central sensor grid positions
	and (b) Allan deviation for a measurement of the quantization field $\sMagneticFieldSensorMagneticQuantizationField$. \textcolor{\redf}{The solid dots represent the mean value for the central 3$\times$3 grid positions, the dashed line represents the inverse-square-root dependence expected for a shot-noise-limited measurement, and the triangle corresponds to the magnetic field resolution $\delta B$ for the total averaging time (see the main text for details).}
	}
	\label{fig:RamseyAllan}
\end{figure}\\
For \textit{in situ} calibration and reduction of long-term drifts, we apply a differential measurement technique: While the quantization field $\sMagneticFieldSensorMagneticQuantizationField$ is activated in every cycle, we switch between cycles with an activated and a deactivated test field $\sMagneticFieldSensorMagneticGradientField$. The data are collected in a pseudorandom way. Throughout this measurement series, on average \num{360(100)} atoms contribute to the Ramsey signal at every sensor pixel for each of these two field configurations (see, e.g., \figref[a]{RamseyAllan}).
\figureref[b]{RamseyAllan} (solid dots) displays the Allan deviation (mean value for central 3$\times$3 grid positions) for the measurement of $\sMagneticFieldSensorMagneticQuantizationField$ as a function of the averaging time following the analysis described in Ref.~\cite{Draganova2014}.
The data point at \SI{3,2e4}{\second} has been derived by dividing the total available dataset into two bins and hence corresponds to half of the averaging time used for determining the resolution and sensitivity analyzed below.
\textcolor{\redf}{
For averaging times less than \SI{3,2e4}{\second} (shaded region), we observe a large variation of the Allan deviation due to the low number of atoms (less than
approximately one on average) detected for each free-precession time step $\sRamseyTimeOfFreePrecession$.
For averaging times of \SI{3,2e4}{\second} and greater, the Allan deviation is significantly reduced and follows the inverse-square-root dependence as a function of averaging time expected for shot-noise-limited measurements (dashed line). This behavior is confirmed by adding datasets for shorter averaging times (open circles), including only values for magnetic field measurements lying within a 3 $\sigma$ confidence interval of the value for long averaging times, and magnetic field resolution $\delta B$ (triangle) determined for an averaging time of \SI{6,4e4}{\second} (see below).}\\
All 270 sites in the sensor array are evaluated in parallel and detailed maps of the Ramsey frequencies for cycles with an activated as well as a deactivated test field are obtained. Both are presented in \figref[]{measurementThreeD}.
%
%
Without the test field, the whole sensor plane shows only slight variations due to the site-specific differential light shift of the tweezers. A nearly constant Ramsey frequency is measured with an average value of $\sMagneticFieldSensorReferenceFrequency=\vMagneticFieldSensorReferenceFrequency$. This corresponds to the preset frequency offset in $\Delta_{12}$ chosen to guarantee an adequate number of Ramsey oscillations within the measurement interval. The activation of the test field leads to a frequency map with a positive gradient along the quantization axis.
There is an intersection of the two planes next to $x=\SI{28}{\micro\meter}$, indicating almost no additional contribution of the test field to the transition energy $\hbar\Delta_{\vStateGroundStateHighMagneticFieldSensorShortForm\vStateGroundStateLowMagneticFieldSensorShortForm}$.
A Ramsey frequency $\sRamseyFrequency<\sMagneticFieldSensorReferenceFrequency$ shows that the test field reduces the effective magnetic field with respect to the reference measurement.
\begin{figure}[t]
	\includegraphics[width=\linewidth]{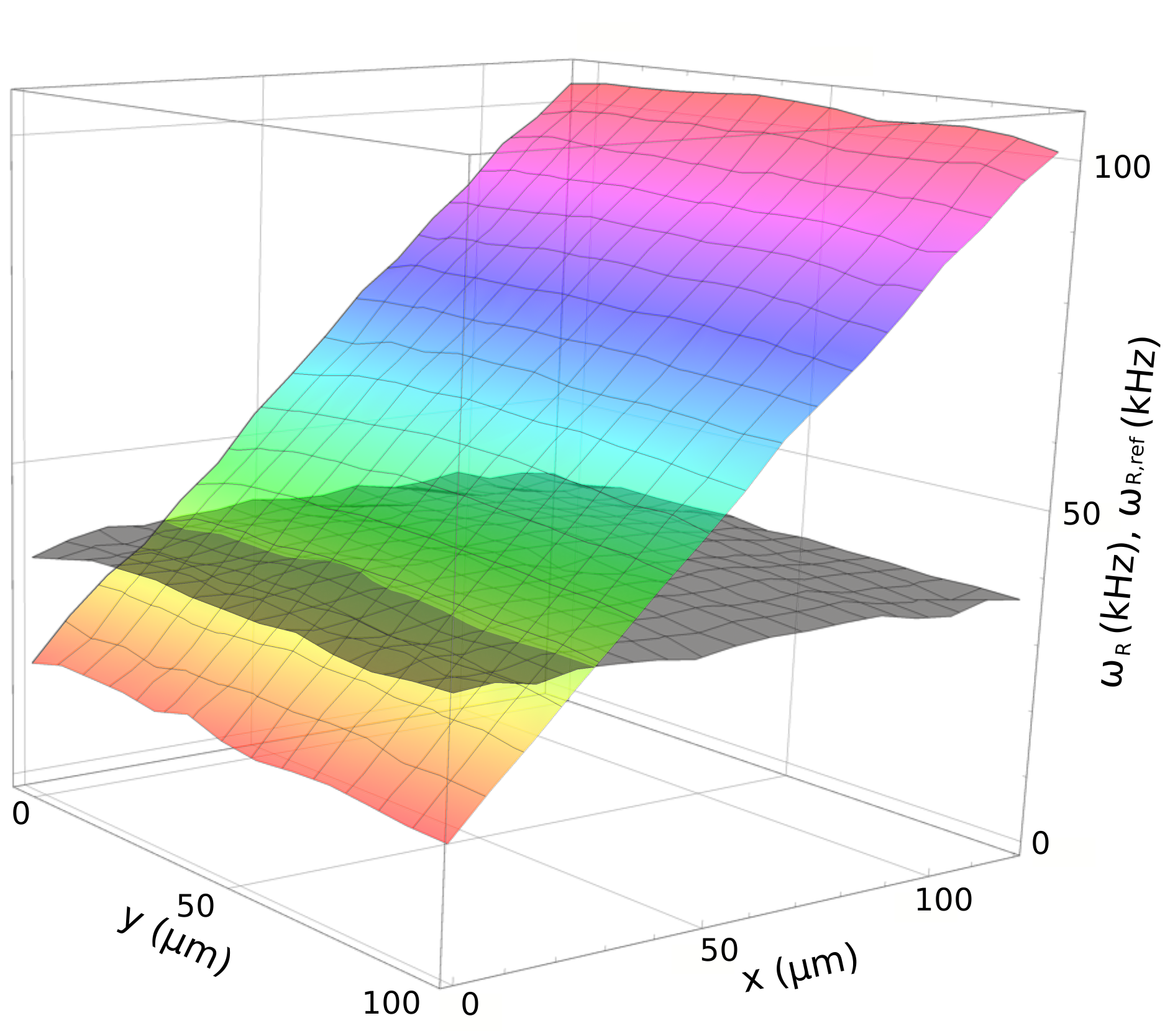}
	\caption{
	    Two-dimensional distribution of the Ramsey frequencies in the sensor plane without (gray) and with (color) an activated magnetic gradient test field $\sMagneticFieldSensorMagneticGradientField$. A linear variation of $\sRamseyFrequency$ along the x direction is observed in the case of an activated test field, while only fluctuations resulting from the local tweezer light shift in $\sMagneticFieldSensorReferenceFrequency$ occur for a deactivated test field.
	    \label{fig:measurementThreeD}
	    }
\end{figure}\\
On the basis of the difference $\Delta\omega_R =\sRamseyFrequency-\sMagneticFieldSensorReferenceFrequency$ between the Ramsey frequencies at each sensor pixel, the shift of the effective magnetic field strength $\Delta B=\partial B / \partial \sMagneticFieldSensorEffectiveDetuningGroundState\times\Delta\omega_R$ 
resulting from the test field can be calculated for the whole sensor plane.
The resulting field map is displayed in \figref[a]{measurement}.
The mean uncertainty of the frequency differences $\Delta\omega_\mathrm{R}$ over the full area amounts to $2\pi\times\SI{0.91(27)}{\kilo\hertz}$, which results in a magnetic field resolution $\delta B$ of $\SI{98(29)}{\nano\tesla}$.
We confirm this value by control measurements of $\sMagneticFieldSensorReferenceFrequency$ for small variations of the calibrated homogeneous quantization field $\sMagneticFieldSensorMagneticQuantizationField$ in a $\SI{3}{}\times\SI{3}{}$ site measurement region (\figref[c]{opticalSetup}).
With increased data rates achieved by repeated assembly of atoms in this subarray \cite{OhldeMello2019}, we are able to distinguish between magnetic fields with minimum steps of approximately $\SI{100}{\nano\tesla}$.
\begin{figure}[t]
	\mysansmath
	\begin{tikzpicture}
	\node[anchor=west,inner sep=0pt] (ResolitonMeasurement) at (0,0) 
	{\includegraphics[]{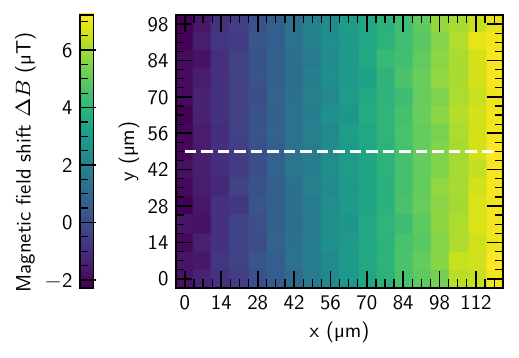}};
	\node[anchor=west,inner sep=0pt] (GradientCrossSection) at (0,-6.1)
	{\includegraphics[]{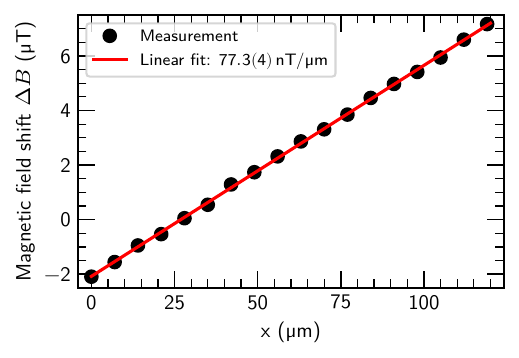}};
	\node[anchor=west] at (0,3.2) {\textbf{(a)}};
	\node[anchor=west] at (0,-2.9) {\textbf{(b)}};
	\end{tikzpicture}
	\caption{Measured shift $\Delta B$ of the effective magnetic field strength. In the two-dimensional map (a), the results of the measurements for each of the 270 pixels equally distributed over the $\SI{119}{\micro\meter}\times\SI{98}{\micro\meter}$ ($\SI{18}{}\times\SI{15}{}$ pixel) sensor plane are given. The magnetic field gradient along the x direction is clearly visible. The data obtained along the dashed line are plotted in (b). From an approximation of a linear function (red) to these data, a gradient of $\SI{77.3(4)}{\nano\tesla/\micro\meter}$ is determined.
		\label{fig:measurement}
		}
\end{figure}\\
\figureref[b]{measurement} gives the data along the dashed line ($y=\SI{49}{\micro\meter}$) in \figref[a]{measurement} and a linear fit (red) to the data.
On the basis of values ranging from $\SI{-2.1(1)}{\micro\tesla}$ at $x=\SI{0}{\micro\meter}$ to $\SI{7.2(1)}{\micro\tesla}$ at $x=\SI{119}{\micro\meter}$, a gradient of $\SI{77.3(4)}{\nano\tesla/\micro\meter}$ is retrieved.
Averaging over the set of gradients determined for the \num{15} rows gives a mean gradient of $\vMagneticFieldSensorGradientMean$ in the sensor area.
Even though the test field exhibits a gradient along the y direction as well, its contribution to the effective magnetic field strength is below the resolution limit $\delta B$ and therefore is not observed when we compare measured data along the y direction for constant x. This is consistent \textcolor{\bluef}{with} a calculation based on the vector characteristics of the magnetic field and the measured gradient along the x direction.\\
%
\textcolor{\bluef}{
We estimate the sensitivity of the
magnetic field measurement following Eq.~(15) in Ref.~\cite{Barry2020}.
Specifically, the overhead time needed for atom preparation and detection has to be taken into account, and is  longer than the free-precession time by orders of magnitude. With the \textcolor{\bluef}{total averaging time of \SI{6.4e4}{\second}, a relative fraction of \num{2.3e-8}} corresponds to the total time of free precession. 
This gives a sensitivity of $\SI{25}{\micro\tesla/\sqrt{\hertz}}$.}\\

\section{discussion}

\textcolor{\bluef}{
Significant straightforward improvements in sensitivity are possible in a next-generation experiment:
(1)} Optimizing the Ramsey sequence by switching to states with maximum $|\sMF|$ increases the susceptibility to magnetic fields by a factor of \num{2.5}, which improves the resolution to $\delta B=\SI{39(12)}{\nano\tesla}$.
(2) Increasing $T_2^*$ will result in a prolongation of the usable Ramsey precession time.
A value of $T_2^*=\SI{5}{\milli\second}$ with optimized laser cooling and larger detuning for trapping and spectroscopy has been achieved in our setup \cite{Lengwenus2010}.
(3) The incorporation of state-of-the-art state preparation will increase the preparation efficiency to $\SI{99}{\percent}$ or greater \cite{Levine2019}.
(4) By applying recent advances in fast nondestructive state detection \cite{Nikolov2023,Chow2023}, repeated assembly of defect-free atom grids \cite{OhldeMello2019}, and modular setups for separating atom preparation from the functional unit for sensing \cite{Pause2023}, a perpetual rate of \textcolor{\bluef}{100} measurement cycles per second is accessible. \textcolor{\bluef}{With these modifications, a sensitivity of $\SI{20}{\nano\tesla/\sqrt{\hertz}}$ is foreseeable in our 2D sensor grid, which will result in a magnetic field resolution below $\SI{0.3}{\nano\tesla}$} after $\SI{1}{\hour}$ of averaging time.\\
So far, spatial resolution has been limited to the geometric constraints of the array. While it enables parallelized detection at all sites of the sensing area and allows the creation of a detailed map of the field distribution at equidistant points, independent control of the specific positions is vital to increase spatial resolution. With use of the movable optical tweezer, the restriction to a periodic geometry can be surpassed and a single-atom scanning probe can be placed at any arbitrary position. Benchmarking this scanning probe with the known properties of the sensor grid, we obtain the same resolution in field strength and sensitivity. The spatial addressability is below than \SI{100}{\nano\meter}, the spatial localization is less than \SI{1}{\micro\meter\squared} for a temperature of \SI{52}{\micro\kelvin} and trap depth $\sArrayTrapDepth$ of $k_\mathrm{B}\times\SI{0.2}{\milli\kelvin}$, and the addressable area amounts to  $\SI{400}{\micro\meter}\times\SI{400}{\micro\meter}$ \cite{Schlosser2020}.\\
\textcolor{\bluef}{
The results reported in this article represent differential measurements that eliminate systematic effects to first order, resulting in the reported resolution $\delta B$ of $\SI{98(29)}{\nano\tesla}$. We did not determine the absolute accuracy of the sensor array. On the basis of the level of control already achieved for the ground-state hyperfine qubit used \cite{Lengwenus2010},
the requirements for obtaining an absolute accuracy comparable to the resolution $\delta B= \SI{98(29)}{\nano\tesla}$ are moderate and we rule out any systematic effects larger than this. 
For a next-generation experiment, we do not expect major limitations by systematic effects for absolute field measurements on the level of 
$\SI{1}{\pico\tesla}$ and a sensitivity of $\SI{1}{\pico\tesla/\sqrt{\hertz}}$ following the numbers reported in Refs.~\cite{Mitchell2020,Budker2007,Vengalattore2007}.
Additional improvements leveraging the full set of techniques applied in cold-atom frequency standards will facilitate concurrent gains in resolution, sensitivity, and absolute accuracy even beyond this level.}\\

\section{summary and outlook}

In summary, in this work we have reported on the application of a tweezer array of individual atoms to quantum sensing. We experimentally demonstrated an optical magnetometer on the basis of a regular 2D sensor grid with micrometer-scale spacing that is built on a microlens-generated scalable optical tweezer architecture.
%
%
\textcolor{\bluef}{
The platform presented was successfully benchmarked through the characterization of a magnetic field with a resolution of \SI{98(29)}{\nano\tesla}, a sensitivity of $\SI{25}{\micro\tesla/\sqrt{\hertz}}$, and a spatial resolution better than  \SI{1}{\micro\meter}.}\\
Already comprising hundreds of grid sites in this work, the scalability of the micro-optical approach enables the near-future implementation of 2D and---based on the Talbot effect---even 3D configurations of thousands of sensor pixels with adaptable geometry \cite{Schlosser2023}. With bias fields and spectroscopy beams oriented along all three spatial coordinates, comprehensive mapping of external fields in a 3D volume can be achieved. 
\textcolor{\bluef}{The required preservation of the quantization axis by a dominating bias field, on the other hand,  might impose limits of applicability, e.g., for magnetically sensitive subjects of analysis. Hence, a suitable working regime can be identified by the maximally allowed bias field strength still excluding deteriorating effects on the sample.\\}
Applications of this sensor grid platform are not limited to magnetometry but can be extended to any external influence on atomic quantum states, e.g.~the susceptibility of Rydberg states to electric fields \cite{Facon2016}. 
%
%
The use of microfabricated optical elements \cite{Birkl2001} facilitates the adaptation of the presented work to integrated setups, where we envision field-deployable atom-array quantum sensors.
%
In addition to single-atom scanning, parallelized individual-atom scanning microscopy for quantum sensing can be achieved by dynamic repositioning of the full sensor grid following the techniques demonstrated in Ref.~\cite{Lengwenus2010}. Including a high-NA microscope objective and ground-state laser cooling has the potential for superresolution imaging \cite{Deist2022} of external fields with spatial resolution on the order of \SI{100}{\nano\meter} or less in a parallelized fashion on the macroscopic sensor grid area.
\textcolor{\bluef}{A direct extension of the presented work is the incorporation of techniques that facilitate optical quantum metrology close to the surface of replaceable samples \cite{Yang2017,Yang2020} which is feasible for materials that tolerate optical power on the milliwatt level at the wavelengths used for trapping and spectroscopy.}
Finally, since our architecture is predestined for the implementation of collective quantum effects \cite{Schlosser2020}, next-generation protocols for quantum sensing exploiting \textcolor{\bluef}{Rydberg cat} states \cite{Dietsche2019}, \textcolor{\bluef}{spin squeezing \cite{Hines2023,Eckner2023,Bornet2023}}, and entanglement \cite{Danilin2018,Ruster2017,Sekatski2020} offer enticing prospects of quantum-enhanced order-of-magnitude improvements, \textcolor{\bluef}{perhaps at the expense of spatial resolution.}
%
\begin{acknowledgments}
	We acknowledge financial support by the Federal Ministry of Education and Research (Grant No. 13N15981), by the Deutsche Forschungsgemeinschaft  [Grants No. BI 647/6-1 and No. BI 647/6-2, Priority Program SPP 1929 (GiRyd)], and by the Open Access Publishing Fund of Technische Universit\"at Darmstadt. D. S. and T. S. contributed equally to this work.
\end{acknowledgments}
%
%
\bibliographystyle{jabbrv_apsrev4-2} 
\bibliography{MagneticFieldSensor} 
\end{document}